\documentclass[aip, jmp, preprint]{revtex4-1}

\usepackage{graphicx}% Include figure files
\usepackage{natbib}
\usepackage{multirow}
\usepackage{amsmath, amssymb}
\usepackage{graphicx}
\usepackage{subfigure}
\usepackage{placeins}
\usepackage{hyperref}
\hypersetup{
    colorlinks=true,
    linkcolor=blue,
    citecolor=blue,
    urlcolor=blue,
    pdfhighlight=/P,
    pdfauthor={Sidharath Jain},
    pdfcreator={LaTeX},
    pdftitle={Broadband, Flat Base, Multibeam, Wide Angle Scan Luneburg Lens Antenna},
    pdfkeywords={Luneburg lens}%,
}

\begin{document}

\title{Flat-Base Broadband Multibeam Luneburg Lens for Wide Angle Scan}% Force line breaks with \\

\author{Sidharath~Jain}
\email[sxj25@psu.edu]{} \affiliation{Electromagnetic Communication Laboratory, Department
of Electrical Engineering, The Pennsylvania State University,
University Park, PA-16802, USA.}

\author{Raj~Mittra}
\email[rajmittra@ieee.org]{} \affiliation{Electromagnetic Communication Laboratory, Department
of Electrical Engineering, The Pennsylvania State University,
University Park, PA-16802, USA.}

\begin{abstract}
In this paper we present the design of a flat-base Luneburg type of lens antenna, designed for wide angle scan. The antenna consists of a $11$-layer lens, fed at its base by a $6$x$6$ array of waveguides. The lens is broadband and has a high aperture efficiency, only $1$ dB below that of a reference aperture antenna with uniform amplitude and phase distributions. Its sidelobe level is $-21$ dB at boresight and $-13$ dB when the scan angle is $64^{\text{o}}$. It shows good performance when compared to the flat Luneburg lens previously reported in the literature, in terms of gain, scan capability, as well as ease of fabrication. It is shown to have the capability of producing multiple beams simultaneously, multiple angle, scan capability. Two different methodologies have been used to design the $6$x$6$ feed array of waveguides for the lens. The first of these utilizes a conventional perfect electric conductor (PEC) waveguide, while the second employs materials for the guided wave region that have a high permittivity at frequencies at which metals become lossy and plasmonic. The performance of the lens has been investigated in this paper for both of these feed array designs.
\end{abstract}

\maketitle

%\begin{IEEEkeywords}
% Flat Lens, Luneburg Lens, Wide Angle Scanning, Broad band Antenna, Spherical Lens.
%\end{IEEEkeywords}

\section{Introduction}
At microwave frequencies, wide-angle scanning is typically achieved either by mechanical means, or by using phased arrays that can be expensive. An attractive alternative is to use a Luneburg lens which offers scan capability over a very wide angular range. One drawback of the original Luneburg lens design, which is spherical, is that it is incompatible with planar feeds or detector arrays that are much more desirable than those that are conformal to the spherical surface. To mitigate this problem, Ma and Cui \cite{cui2010} have employed the Transformation Optics (TO) algorithm to transform the spherical shape of the Luneburg lens into a rectangular box, at least partially. However, Smith et al. \cite{smith2012} have pointed out that performance of this lens is not very satisfactory for large scan angles, say beyond $30^{\text{o}}$, and the design is polarization dependent. This is because flattening a part of the spherical lens restricts its field of view (FOV). And if the FOV angle is increased, the material parameters dictated by the TO algorithm become highly anisotropic and less than $1$, both of which are undesirable.

 Returning to the TO algorithm, it is well known that it is based on geometry transformation, which preserves the field variation when we map the physical system into a virtual one, or vice versa \cite{Leon06,pendry06,pendry08}. The TO does provide us the information on the material parameters ($\epsilon_r$, $\mu_r$) in the physical system from the knowledge of those in the virtual system. The caveat is, though, that sometimes these values, dictated by the TO, can be difficult to realize in practice because they may either be less than $1$, or much greater than $1$. The quasi-conformal transformation optics (QCTO) has been proposed as a way to reduce the anisotropy of the medium, and to set $\mu_r=1$, albeit at a cost, since such an approximation often degrades the performance. Also, not unexpectedly, the QCTO approximation breaks down for 3D geometries, and this prompts one to turn to a 2D quasi-conformal type of transformation, as an approximation; however, this restricts its application to a single polarization, which is obviously undesirable \cite{smith2012}. Therefore, while designing practical lens using TO the regions with $\epsilon_r < 1$ are replaced with free space which limits the scan performance of the lens. Thus, the material has to be assumed to be nonmagnetic ($\mu_{rx,ry,rz}=1$), isotropic ($\epsilon_x=\epsilon_y=\epsilon_z$) and non-metallic (materials whose $\epsilon_r>1$) for any practical designs to be feasible \cite{smith2012}.

 Bosiljevac et al. \cite{stefano2012} have proposed a flat metasurface Luneburg lens antenna, which is shown to have slightly higher sidelobe levels compared to those of a flat dielectric Luneburg lens. However, this lens is narrowband and, furthermore, its gain reduces rapidly as we increase the scan angle because of impedance mismatch problems. Thus, the Luneburg lens designed by using this meta-surface technique loses its wide-angle scan capability, for which it was originally designed. Also, the meta-surface technique only allows one to develop lenses which can scan from boresight either along the azimuth or elevation, which makes them unsuitable for many applications that require scan capabilities both along the azimuth as well as elevation.

 In this paper we propose a novel design for the Luneburg lens, which preserves the broadband, high gain, wide FOV coverage and low sidelobe levels of the original Luneburg lens design. It is capable of simultaneously transmitting multiple beams along designated azimuth or elevation angles. The proposed design has a flat base (see Fig.\ref{Fig2}) and, hence, it is compatible with planar type of feeds, which is highly desirable for many applications. Two different approaches to realizing the waveguide array feed have been demonstrated. The first one of these is based on using a perfect electric conductor (PEC) waveguide, while the other, utilizes a waveguide with walls made with high $\epsilon_r$ materials separated by a small air gap. The proposed antenna is designed to scan at different angles, simply by exciting different portions of the waveguide array feed. Furthermore, the lens is broadband since it uses conventional isotropic materials, as opposed to anisotropic metamaterials. The lower edge of the frequency band is limited by the size of the feed waveguide which, in turn, determines its cut-off frequency, while the upper edge of the frequency band is determined by the level of discretization used to fabricate the lens. Also, the antenna has relatively low sidelobes (below $-20$ dB), as compared to conventional array antennas, where the side-lobe suppression can be a challenge and can cause false alarms in RADAR applications if their level is not reduced.

\begin{figure}[htp]
\centering
\includegraphics[width=8cm]{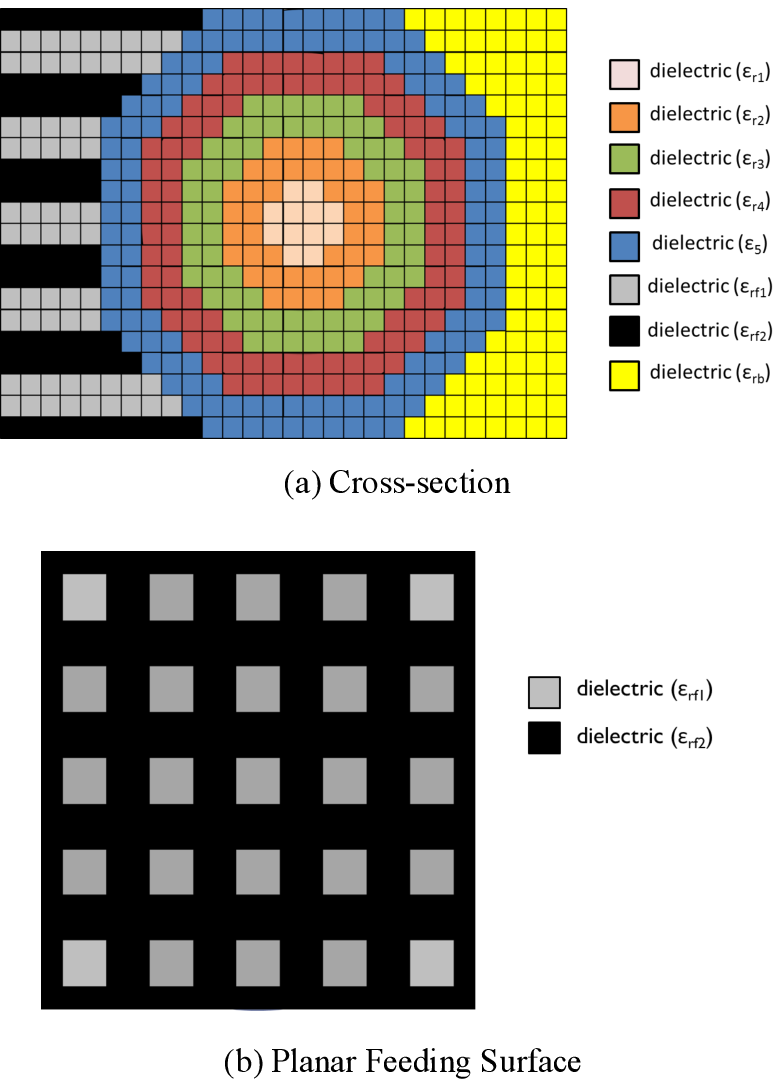}
\caption{All Angle Scan Luneburg Lens Design (a) Discretized Cross Section (illustration) (b) Feeding Plane}
\label{Fig0} % caption for the whole figure
\end{figure}

\begin{figure}[h]
\begin{center}
\noindent
  \includegraphics[width=7cm]{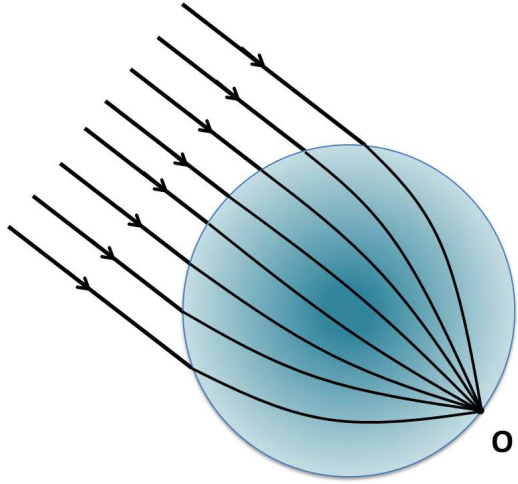}
  \caption{Luneburg Lens Principle}\label{Fig2}
\end{center}
\end{figure}

\begin{figure}[htp]
\centering
\includegraphics[width=6cm]{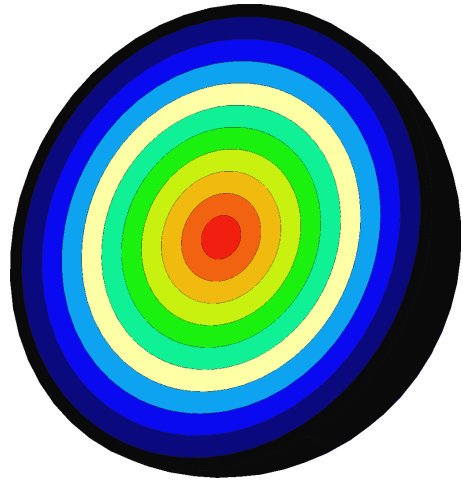}
\caption{Cross Section of the designed Luneburg Lens}
\label{Fig1} % caption for subfigure a
\end{figure}

\begin{figure}[htp]
\centering
\includegraphics[width=8cm]{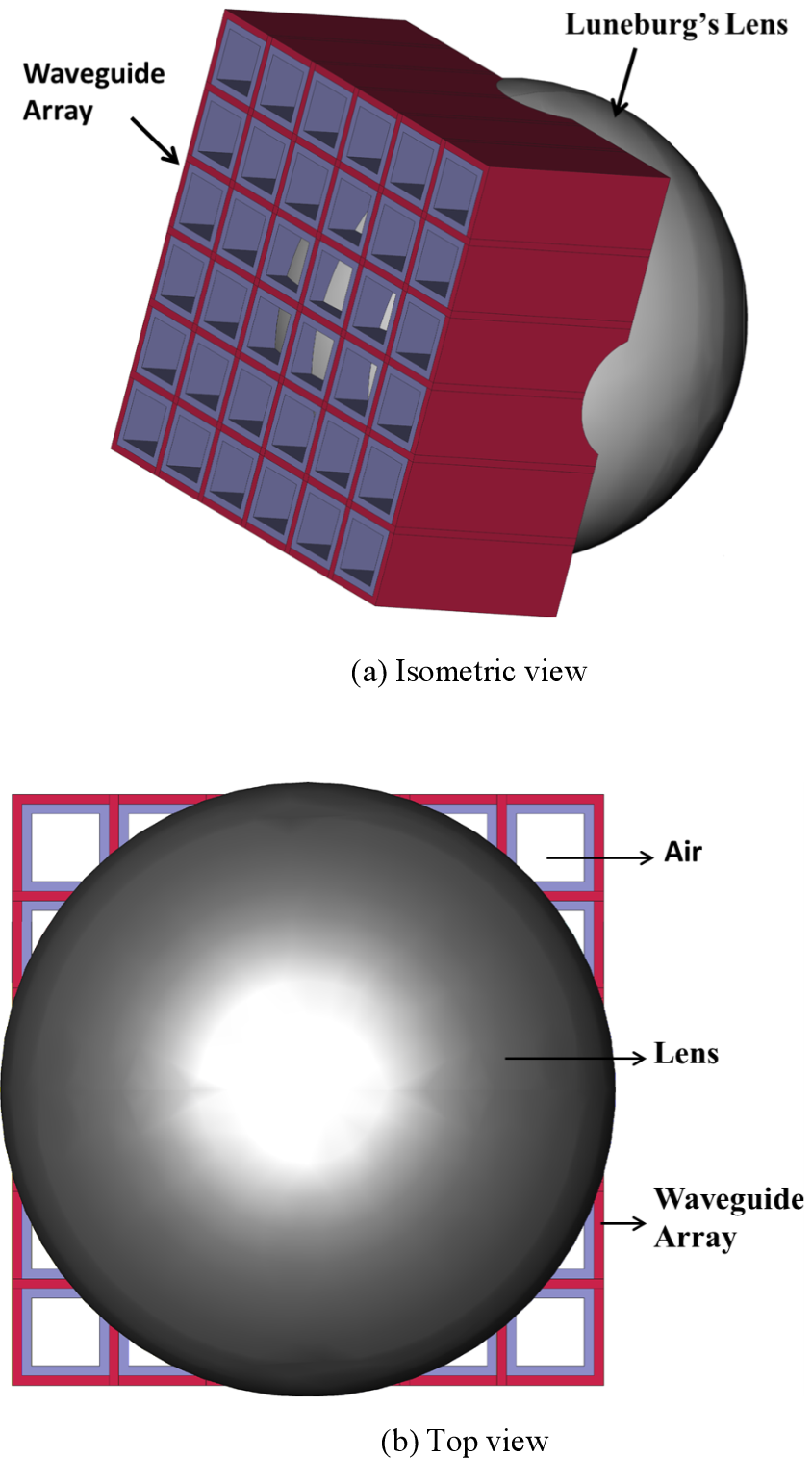}
\caption{Designed Large Angle Scan Luneburg Lens}
\label{Fig3} % caption for the whole figure
\end{figure}

\begin{figure}[h]
\begin{center}
  \includegraphics[width=8cm]{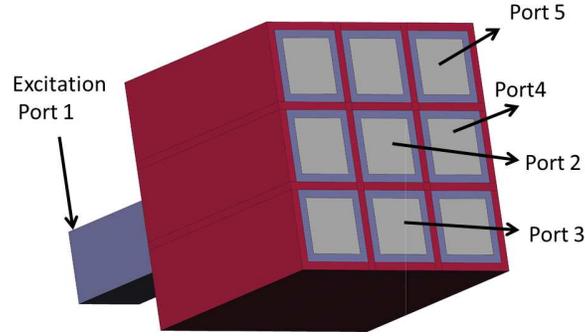}
  \caption{Waveguide Array Design}\label{Fig4}
\end{center}
\end{figure}

\begin{figure}[h]
\begin{center}
  \includegraphics[width=8cm]{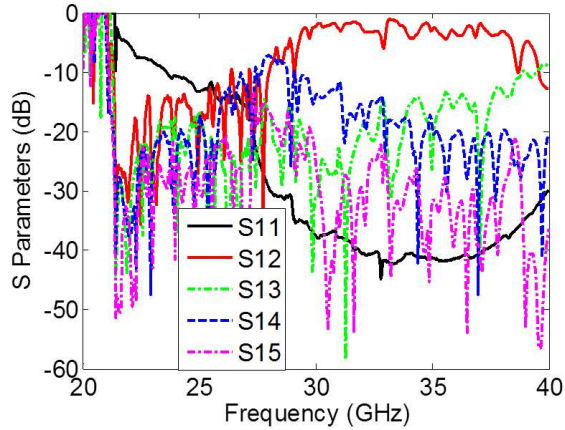}
  \caption{S-parameters for the five ports in Fig.\ref{Fig4}}\label{Fig5}
\end{center}
\end{figure}

\begin{figure}[h]
\begin{center}
  \includegraphics[width=8cm]{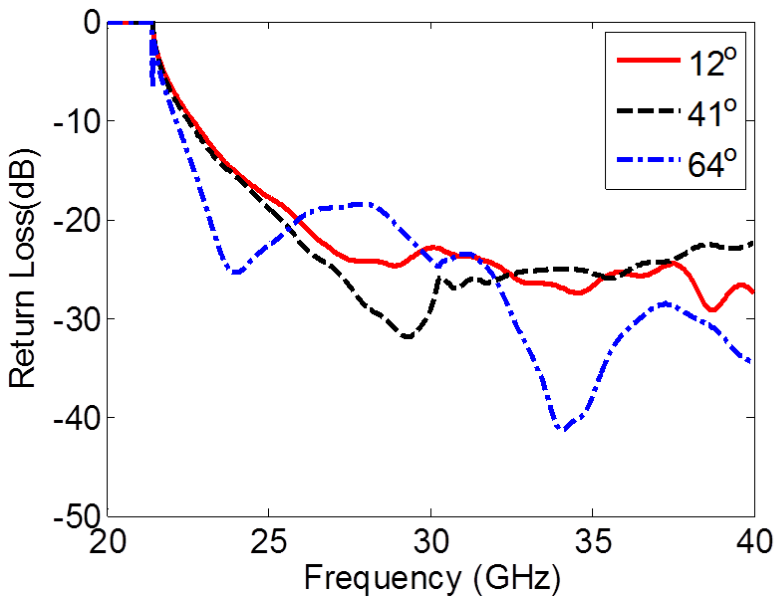}
  \caption{Return Loss at the excitation port when ports shown in red in Figs. \ref{Fig8}, \ref{Fig9} and \ref{Fig10}, respectively are excited with the $TE_{10}$ }\label{Fig6}
\end{center}
\end{figure}

\section{Lens Antenna Design}

Fig.\ref{Fig0} illustrates the underlying principle behind the design of the Luneburg lens antenna for wide-angle scan. The lens is designed to focus a plane wave, arriving from an arbitrary direction, at a point which is located diametrically opposite to that of the incident side, as shown in Fig.\ref{Fig2}. Luneburg has shown that the problem of finding the $\epsilon_r$ of the lens can be formulated in terms of an integral equation \cite{Luneburg}, whose solution provides us the required material parameters of the lens. They are given by:

\begin{figure}[htp]
\centering
\includegraphics[width=8cm]{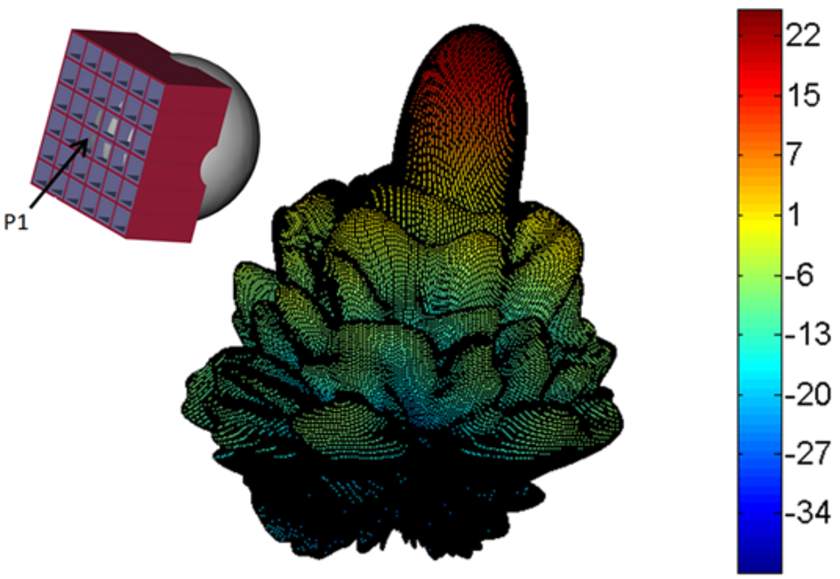}
\caption{Radiation Pattern of the Luneburg Lens Antenna when a $TE_{10}$ mode is excited in port P1.}
\label{Fig8} % caption for the whole figure
\end{figure}

\begin{figure}[h]
\centering
\includegraphics[width=8cm]{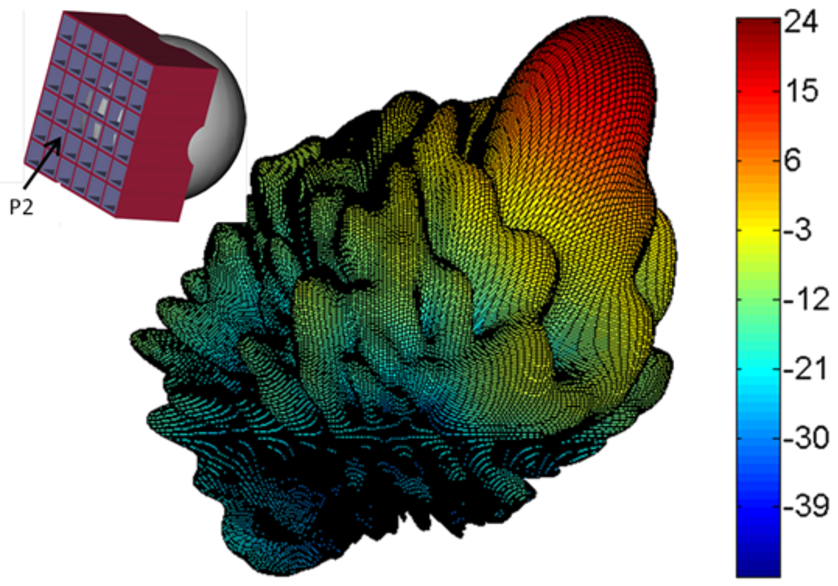}
\caption{Radiation Pattern of the Luneburg Lens Antenna when a $TE_{10}$ mode is excited in port P2.}
\label{Fig9} % caption for the whole figure
\end{figure}

\begin{table*}[ht]
\begin{center}
\renewcommand{\arraystretch}{1.4}
\caption{Material Parameters of the Spherical Luneburg's Lens in Fig.\ref{Fig1}} \label{Table1}
\begin{tabular}{c c c c c c c c c c c}
 \hline
 $\bf{\epsilon_{r1}}$&	 $\bf{\epsilon_{r2}}$&$\bf{\epsilon_{r3}}$&$\bf{\epsilon_{r4}}$&$\bf{\epsilon_{r5}}$&$\bf{\epsilon_{r6}}$&$\bf{\epsilon_{r7}}$&$\bf{\epsilon_{r8}}$&$\bf{\epsilon_{r9}}$&$\bf{\epsilon_{r10}}$&$\bf{\epsilon_{r11}}$\\
   \hline
 2.0&	1.96&	1.92&	1.86&	1.78&	1.68&	1.56&	1.43&	1.28&	1.11&	1.05\\
 \hline
\end{tabular}
\end{center}
\end{table*}

\begin{table*}[!ht]
\centering
\renewcommand{\arraystretch}{1.4}
    \caption{Comparison of Gain (in dBi) of three different designs of the Luneburg Lens Antenna shown in Fig.\ref{Fig3} as a function of frequency for a scan angle of $41$ degrees.}
    \label{Table2}
     \begin{tabular}{c c c c c }
      \hline
      & \multicolumn{4}{c}{\bf{Gain (dBi)}} \\
      \hline
      {\bf {Frequency (GHz)}} & {\bf{Constant}} &{\bf {PEC with Air Gap}} &{\bf {$\epsilon_r$=50 with Air Gap}} &{$\bf {PEC}$}\\
     \hline
     24  & 24.06 & 21.73 & 12.41 & 21.89\\
     \hline
     28  & 25.40 & 22.86 & 20.00 & 23.30\\
     \hline
     30 & 25.99 & 23.42 & 21.88 & 23.75\\
     \hline
     36 & 27.58 & 23.61 & 21.85 & 24.14\\
     \hline
     40 & 28.50 & 24.05 & 20.48 & 24.46\\
     \hline
    \end{tabular}
   % \end{small}
\end{table*}

\begin{equation}
\label{eq:eq1}
\epsilon_r=2-\Big(\frac{r}{R}\Big)^2
\end{equation}
where $r$ is the distance from the center of the lens, and $R$ is the radius of the lens. It is evident from Fig.\ref{Fig2}, that if we place a feed whose phase center is located on the surface of the lens, it would transform the spherical wavefront emanating from the phase center into a planar one. If we wish to scan the beam, obviously we would have to rotate the feed around the surface of the lens, which is not the most convenient thing to do; instead, it would be considerably more desirable to place the feed array on a flat surface. With this in mind, we propose a design, shown in Fig.\ref{Fig2}, of a lens with a spherical profile, which preserves the salutary features of the conventional Luneburg lens, but is much more convenient to feed since it has a flat-base design. It consists of $11$ layers and has a diameter of $63.5$ mm. The first ten layers from the center have a thickness of $3$ mm each and the last layer is $1.75$ mm thick. The dielectric constants of the layers vary depending on their distance from the center according to \eqref{eq:eq1} and are listed in Table \ref{Table1}. The materials required to fabricate the Luneburg lens can be realized by using a number of techniques which have been developed recently for synthesizing artificial dielectrics \cite{eff_medium,Zhang,Vardax,Smith2010,3Dprinting,3Dprinting1}. The performance of the spherical Luneburg lens fabricated by using an alternative technique for realizing the material properties of the artificial dielectric has been reported by Smith \cite{smith2012}. Theoretically, there is no limit on the upper frequency of operation. If the material has been realized by drilling holes in the dielectric, then the upper frequency for this type of approach is determined by the breakdown of effective medium theory which occurs when the resonant length of the holes becomes comparable to the wavelength. The losses of the materials designed using effective medium theory are comparable to those in typical substrate materials such as FR4 \cite{smith2009}.

\begin{figure}[h]
\centering
\includegraphics[width=8cm]{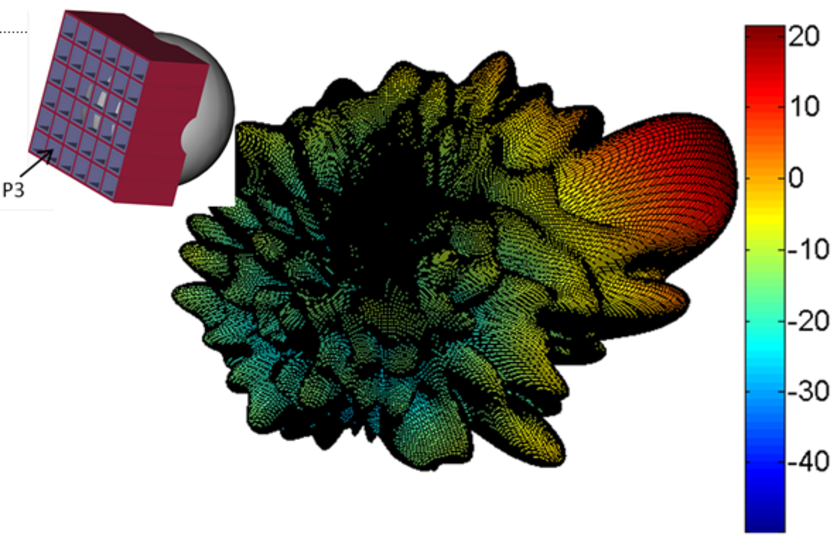}
\caption{Radiation Pattern of the Luneburg Lens Antenna when a $TE_{10}$ mode is excited in port P3}
\label{Fig10} % caption for the whole figure
\end{figure}

\begin{figure}[h]
\centering
\includegraphics[width=8cm]{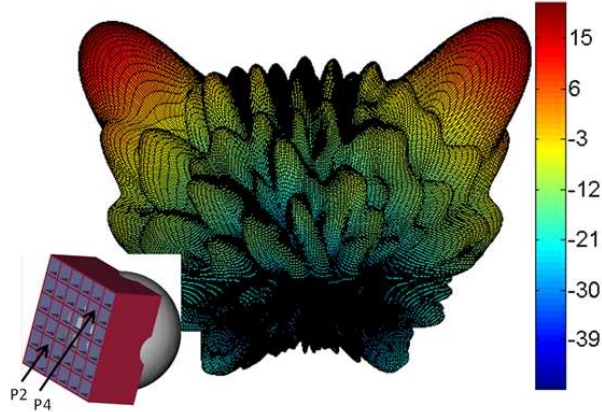}
\caption{Radiation Pattern of the Luneburg Lens Antenna when $TE_{10}$ mode is excited in ports P2 and P4 demonstrating simultaneous multiple angle scan capability.}
\label{Fig11} % caption for the whole figure
\end{figure}

\begin{figure}[h]
\centering
\includegraphics[width=8cm]{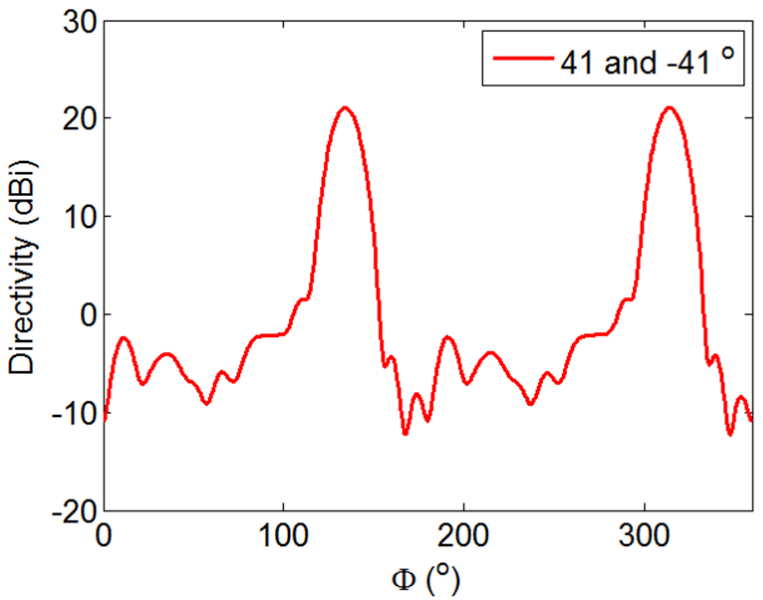}
 \caption{Cross Section of radiation pattern in Fig.\ref{Fig11} for $\theta$ = $41^o$ demonstrating simultaneous multiple angle scan capability.}
\label{Fig15} % caption for the whole figure
\end{figure}
 The proposed flat-base Luneburg  lens is fed by using a $6$x$6$ waveguide array. A beam pointing at the desired angle is generated by feeding one of these waveguides. The beam can be scanned, up to $72^{\text{o}}$ from boresight in both azimuth and elevation, by exciting one of the $36$ waveguides. The size of each waveguide comprising the feed array is chosen such that its cut-off frequency is just below the lowest frequency of operation of the desired antenna. Each waveguide comprising the array has a square cross-section with side of length $7$mm, and is separated from the adjacent waveguides by a $3$mm thick wall. In order to design a feed array for the Luneburg Lens it is important to ensure that the coupling between the adjacent waveguides be negligible, because any type of leakage deteriorates the performance of the lens, and choosing the waveguide walls to be PEC would satisfy the above criterion. However, this would limit the applications of the design up to a few hundred GHz, as most metals exhibit plasmonic properties above these frequencies. Therefore, an all-dielectric design of the waveguide array feed was also investigated for such high frequencies, for which the walls were designed by using $3$mm thick dielectric material with a very high $\epsilon_r=50$. But this was unable to bring the coupling between the adjacent waveguides below the desired $-10$ dB level. To circumvent this problem, the $3$mm thick wall was redesigned by using a $3$-layer sandwich consisting of a $1$mm thick dielectric with $\epsilon_r=50$ on either side, separated by an airgap or foam reduced the coupling below the desired $-10$ dB level. A $3$x$3$ waveguide array, with waveguide walls made of $1$mm thick dielectric material, and with $\epsilon_r=50$ and a $1$mm air gap between two adjacent waveguides (see Fig. \ref{Fig4}) was used to determine the level of energy coupled into the neighboring waveguides. The S-parameters for this array are shown in Fig.\ref{Fig5}. The Figure shows that the coupling between the adjacent waveguides is well below $-10$ dB, and that the direct coupling from port $1$ to port $2$ is close to $-1$ dB for frequencies above the cut-off frequency of the waveguide. The waveguide apertures are parallel to a plane tangential to the lens, and are located at a distance of $2.25$mm from it. To see whether there is a significant impedance mismatch at the point where the waveguide meets the lens surface (see  Fig.\ref{Fig5})  we compute the return loss when three different waveguides shown in Figs.\ref{Fig8}, \ref{Fig9} and \ref{Fig10} excite three beams at $13^o$, $41^o$ and $64^o$, respectively. The return losses for all the three cases are well below the $-20$ dB level, indicating that the impedance match is good. The phase correction due to the curved surface of the lens must be applied when exciting the different waveguides for a particular application. All simulations have been carried out by using a commercial FDTD code. A waveport is used to excite the dominant $TE_{10}$ mode in the waveguide which can be done in practice by placing a dipole at the center of the feeding face of the waveguide. Also, the antenna can handle an arbitrary polarization. Since metals such as copper become lossy above $30$ GHz, gold and silver should be used. At still higher frequencies one can use the all-dielectric design discussed above, whose performance is comparable to that of the PEC when used for the waveguide region.

\begin{figure}[htp]
\centering
\includegraphics[width=7cm]{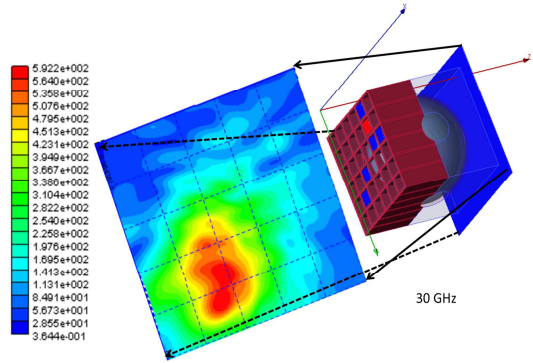}
\caption{Electric field amplitude at the aperture plane of the lens shown in blue when one of the wave guides shown in red is excited.}
\label{Fig7} % caption for the whole figure
\end{figure}
\begin{figure}[htp]
\centering
\includegraphics[width=7cm]{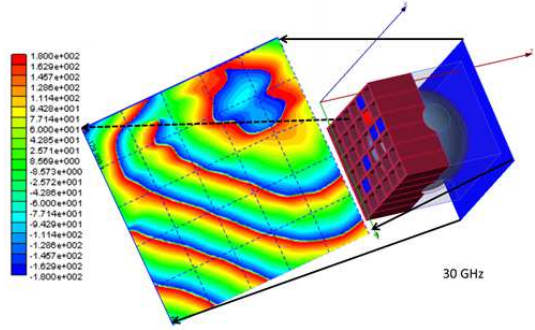}
\caption{Phase of electric field at the aperture plane of the lens shown in blue when one of the wave guides shown in red is excited.}
\label{Fig7a} % caption for the whole figure
\end{figure}

\begin{figure}[h]
\begin{center}
\includegraphics[width=3.0in]{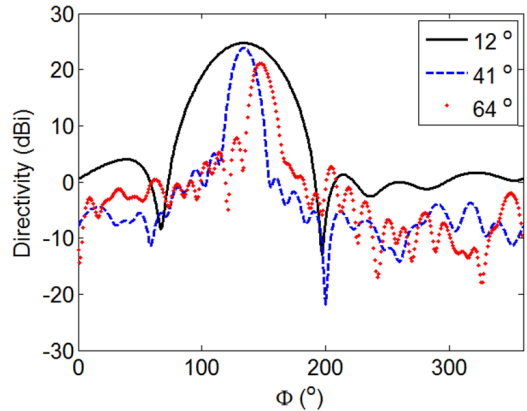}
\caption{Directivity of the Luneburg Lens Antenna as a function of $\phi$ for different scan angles for $\theta$ =$12^o$, $41^o$ and $64^o$ cut.} \label{Fig12} \vspace{-.5cm}
\end{center}
\end{figure}

\section{Results}
\begin{figure}[h]
\begin{center}
\includegraphics[width=3.0in]{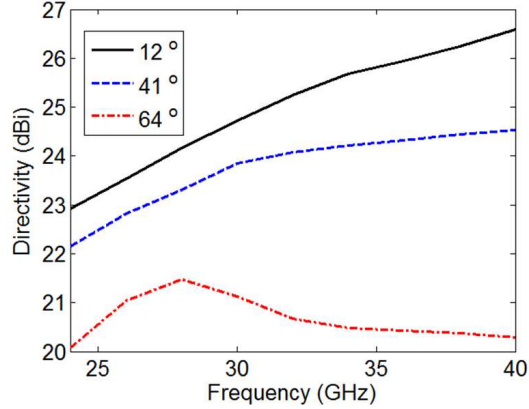}
\caption{Maximum directivity of the Luneburg Lens Antenna as a function of frequency for different scan angles for $\theta$ =$12^o$, $41^o$ and $64^o$ cut.} \label{Fig13} \vspace{-.5cm}
\end{center}
\end{figure}

Figs.\ref{Fig8}, \ref{Fig9} and \ref{Fig10} show the far field patterns of the proposed antenna when three different waveguides denoted by P1, P2 and P3 are excited to generate beams at $13^o$, $41^o$ and $64^o$, respectively. Figs.\ref{Fig7} and \ref{Fig7a} show the amplitude and phase variations in the exit aperture plane when the waveguide shown in red is excited to point a beam at an angle of $41^o$ from the zenith, either in the azimuth or in the elevation plane. The amplitude has a maximum field in a region in the aperture plane, which is diametrically opposite to the one of the feed, as shown in Fig.\ref{Fig7}. Also, the phase distribution in Fig.\ref{Fig7a} shows that the spherical wavefront is transformed into a planar wavefront. A maximum scan angle of $64^o$ has been achieved, using this structure and the scan performances in terms of directivity variations are shown in Figs.\ref{Fig12} and \ref{Fig13}, as a function of scan angle $\phi$ and frequency, respectively.

  Fig.\ref{Fig12} plots the directivity (in dBi) for scan angles of $12^{\text{o}}$, $41^{\text{o}}$ and $64^{\text{o}}$ . The directivities of the waveguide-fed Luneburg lens antenna for scan angles of $12^{\text{o}}$, $41^{\text{o}}$ and $64^{\text{o}}$ are $24.73$, $23.84$ and $21.12$ dBi, respectively. These figures are just $1.26$, $2.15$ and $4.87$ dB below, respectively, for these three scan angles in comparison to that for a uniform aperture distribution, which has a directivity of $25.99$ dBi. In comparison to the proposed lens design, the directivity of the flattened TO lens \cite{cui2010} (aperture size $5.4\lambda$) is $21$ dBi for the scan angle of $45^{\text{o}}$, which is $3.59$ dB below that of an aperture with a uniform field distribution ($24.59$ dBi). Also, as shown in Fig.\ref{Fig13}, the first sidelobe level for the proposed design, normalized w.r.t. the mainlobe, is less than $-20$ dB for small angles and it rises to $-13$ dB level when the scan angle is increased to $64^{\text{o}}$. In contrast to this, the sidelobe level for the metasurface Luneburg lens antenna \cite{stefano2012} is $-6.5$ dB, even at boresight, and the level progressively deteriorates when the scan angle is increased.

  It is interesting to note that the proposed antenna can simultaneously achieve directed beams at multiple angles, if desired. Figs.\ref{Fig11} and \ref{Fig15} show the radiation pattern of the lens antenna when two of the waveguides, marked as P2 and P4, are excited simultaneously. The directivity of the antenna along both the scan angles i.e. $-41^o$ and $41^o$ from the zenith is nearly $22$ dBi with sidelobe levels $-23$ dB w.r.t. the main lobe that are even lower than that for the single angle scan.

The $3$ dB beamwidth of the lens is approximately $16^{\text{o}}$ for a scan angle of $64^{\text{o}}$, as shown in Fig.\ref{Fig12}. We note that the antenna has a very wide field of view up to $72^{\text{o}}$ from the zenith, in both azimuth and elevation, and that it exhibits similar performance characteristics for both polarizations. Furthermore, the structure is broadband, since it has no obvious limitations on the maximum frequency of operation, as there is when metamaterials are used to fabricate the lens. The lower end of its usable frequency is determined by the cutoff frequency of the waveguide, which is $21.4$ GHz for the design shown in Fig.\ref{Fig2}. Fig.\ref{Fig13} shows that maximum gain of the Luneburg lens antenna is high across a wide frequency band and it increases with frequency, as expected.

The scan angle is slightly less than the designed one because the feeding waveguides are not orthogonal to the lens which introduces a initial phase difference at the input port, and which results in the rotation of the beam by a few degrees. Also, as we can see in the Fig.\ref{Fig7}, the beamwidth is higher for normal incidence but decreases as the incidence angle is increased. This is because the aperture size of the excitation port increases from $A$ to nearly $2A$ so  the beam width decreases from $1/A$ to $1/2A$ as is predicted by the Fourier transform relationship between the aperture size and the far field pattern.

Table \ref{Table1} shows a comparison of the gain as a function of frequency for three different feed designs. The maximum gain for the case with constant aperture field distribution is calculated by using the relation:

\begin{equation}
G_{max}= \frac{4 \pi A_e \eta}{\lambda^2}
\end{equation}

where $A_e$ is the effective aperture area and $\eta$ is the aperture efficiency which is chosen to be $1$ for $G_{max}$. The feed structure utilizing the PEC waveguide array, with no gap between two adjacent guide walls has the best performance, followed by the PEC waveguide array with air gap and the dielectric waveguide array with air gap, in that order.

\section{Conclusion}

In this paper, the design of a spherical Luneburg lens with a planar feed structure, which is easy to fabricate and is compatible with the traditional feed array designs, has been presented. Its performance has been shown to be superior to the flattened Luneburg lens designed by using the TO methodology, both in terms of gain (by $1$ dB) as well as angular scan capability (by $34^o$). Also, the antenna has the potential to replace the existing array antenna technology used in radar, satellite communication and other beam scanning applications for a number of reasons: (i)it is easy to design and to fabricate; (ii) it has a wide-angle scan capability; (iii) broad bandwidth; (iv) comparatively low side-lobe levels; and (v) multiple angle beampointing capability, which is particularly desirable in some applications.

\bibliographystyle{aipnum4-1}
%\bibliography{luneburg_jap}
%\nocite{*}
%merlin.mbs aipnum4-1.bst 2010-07-25 4.21a (PWD, AO, DPC) hacked
%Control: key (0)
%Control: author (8) initials jnrlst
%Control: editor formatted (1) identically to author
%Control: production of article title (-1) disabled
%Control: page (0) single
%Control: year (1) truncated
%Control: production of eprint (0) enabled
%

\end{document}